# Controlled-Not Quantum Logic Gate in Two Strongly Coupled Semiconductor Charge Qubits

Hai-Ou Li[1,3,*], Gang Cao[1,3,*], Guo-Dong Yu[2,3,*], Ming Xiao[2,3,a)], Guang-Can Guo[1,3], Hong-Wen Jiang[4], and Guo-Ping Guo[1,3,b)]

[1] *Key Laboratory of Quantum Information, CAS, University of Science and Technology of China, Hefei, Anhui 230026, China*

[2] *Department of Optics and Optical Engineering, University of Science and Technology of China, Hefei, Anhui 230026, China*

[3] *Synergetic Innovation Center of Quantum Information & Quantum Physics, University of Science and Technology of China, Hefei, Anhui 230026, China*

[4] *Department of Physics and Astronomy, University of California, Los Angeles, CA 90095, USA*

[*] *These authors contributed equally to this work.*

[a),b)] *Authors to whom correspondence should be addressed. Electronic mail addresses:*
<u>maaxiao@ustc.edu.cn</u>; <u>gpguo@ustc.edu.cn</u>

**Abstract**

A crucial requirement for scalable quantum-information processing is the realization of multiple-qubit quantum gates. Universal multiple-qubit gates can be implemented by a set of universal single qubit gates and any one kind of two-qubit gate, such as a controlled-NOT (CNOT) gate. Semiconductor quantum dot (QD) qubits are a leading approach for the physical implementation of quantum computation, due to their potential for large-scale integration. Two-qubit gate operations have been so far only demonstrated in individual electron spin-based quantum dot systems. Due to the relatively short de-coherence time, charge qubits in quantum dots are generally considered to be inferior for going beyond single qubit level. Here, we demonstrate the benchmarking CNOT gate in two capacitively coupled charge qubits, each consisting of an electron confined in a GaAs/AlGaAs double quantum dot (DQD). Owing to the strong inter-qubit coupling strength, gate operations with a clock speed up to 5GHz has been realized. A processing tomography shows encouragingly that the universal two-qubit gate operations have comparable fidelities to that of spin-based two-qubit gates. Our results suggest that semiconductor charge qubits have a considerable potential for scalable quantum computing and may stimulate the use of long-range Coulomb interaction for coherent quantum control in other devices.

**Introduction**

Semiconductor quantum dots, hailed for their potential scalability, are outstanding candidates for solid state-based quantum information processing [1-3]. Qubits, encoded by the charge occupancy of a single electron in a double quantum dot, have attracted considerable attentions [4-9] for number of reasons. First, speed of gate operation is primarily determined by the inter-dot tunneling rate which can be made to be extremely fast. Second, initialization, manipulation, and read-out, are all intuitively simple in this all electrical approach. Furthermore, since the exclusive use of charge degree of freedom for computation is compatible with the mainstream information processing technology, one may take the advantage of the wealth of the semiconductor infrastructures for scaling up to large-scale quantum circuits.

One of the basic building blocks of universal quantum computation is a two-qubit gate. However, the implementation of a two-qubit gate operation in QD charge qubits has not been demonstrated to date, largely due to the technical challenges of archiving strong coupling between qubits and the ability to control gate pulses in the nanosecond time scales [10-13].

In this letter, we report the coherent manipulation of a capacitively coupled qubit pair. We achieve a strong electrostatic dipole coupling between two charge qubits. The large coupling energy enables us to completely and coherently turn on/off the Rabi oscillations of one qubit by pulse-driving the charge on the other qubit. A CNOT operation is demonstrated based on this effect [14-19]. In addition, we combined this CNOT gate and universal single qubit gates by using Landau-Zener interferences, to show the feasibility of achieving arbitrary two-qubit gates in this system.

Our results also demonstrate that the fidelity of two-qubit operations for a QD charge qubit can be just as high as that of spin-based semiconductor qubits [20-22]. For charge qubits, with a sufficiently large coupling energy, the fidelity of two-qubit operations is only limited by the fidelity of the single qubit. Thus, with the reduction of decoherence rate of single-qubit using a more sophisticated double quantum dot dispersion [8], the prospect of semiconductor charge qubits for scalable quantum computation can be considerably improved.

**1 Strong Inter-qubit Coupling**

Fig. 1(a) depicts our two-qubit system consisting of two coupled DQDs and two quantum point contacts (QPCs). The Hamiltonian of this system is as follows:

$$H_{2q} = \frac{\varepsilon_U \sigma_z + \Delta_U \sigma_x}{2} \otimes I + I \otimes \frac{\varepsilon_L \sigma_z + \Delta_L \sigma_x}{2} + J \frac{I - \sigma_z}{2} \otimes \frac{I - \sigma_z}{2} \qquad (1)$$

Here, $\varepsilon_U$ ($\varepsilon_L$) is the energy detuning, $\Delta_U = 2t_U$ ($\Delta_L = 2t_L$) is twice the inter-dot tunneling rate for the upper (lower) DQD, $\sigma_x$ and $\sigma_z$ are the Pauli matrixes, $I$ is the unitary matrix, and $J$ is the inter-qubit coupling energy. We denote the four eigenstates of the above Hamiltonian by |00>, |10>, |01> and |11>, and we will discuss gate operation and state evolution in the basis defined by these eigenstates [23].

The inter-qubit coupling energy $J$ originates from the Coulomb repulsion between an electron in the upper DQD and another electron in the lower DQD. When the two electrons are closest to each other, the Coulomb interaction energy is higher, by an amount defined as $J$, than it is when they are furthest apart from each other [10-13]. This is illustrated in Fig. 1(b): the abrupt energy shift from state |00> to state |11> is given by $J$. We will see that a sufficiently large $J$ is the key to controlling the coherent rotations of one qubit by manipulating the state of the other qubit and is, therefore, also the key to realizing two-qubit gates such as CNOT gates.

In our experiment, we are able to achieve very high $J$ ($\approx$ 29.0 GHz) compared with other characteristic parameters: $\Delta_U \approx$ 6.2 GHz and $\Delta_L \approx$ 6.0 GHz. Fig. 1(c) presents the differential current of the upper QPC, and Fig. 1(d) presents that of the lower QPC. Therefore, Fig. 1(c) records only the response to the upper detuning, $\varepsilon_U$, and Fig. 1(d) records only the response to the lower detuning, $\varepsilon_L$. The two figures together constitute a complete description of Fig. 1(b). We find that $J$ is equal to approximately 119 µeV ($\approx$ 29.0 GHz) using the energy-voltage conversion factor obtained from transport measurements of 30 µeV/mV. We can deliberately tune the voltages on the two horizontal gates $H_1$ and $H_2$ to maximize the inter-qubit coupling energy $J$ and simultaneously suppress the direct inter-qubit tunneling.

Now, we apply a rectangular voltage pulse to one of the upper qubit's gates, $U_1$. We initialize both the upper and lower qubits in state |0>, i.e., $-\varepsilon_{U,L} \gg J \gg \Delta_{U,L} > 0$. Under these conditions, the two qubits are nearly uncorrelated, and only the upper qubit is affected by the voltage pulse. We choose the pulse amplitude such that it will drive the upper qubit exactly to its balance point. By sweeping the pulse width, we induce Rabi oscillations in the upper qubit [24]; the qubit oscillates between states |0> and |1>, with a probability in each state as a cosine function of the pulse width $W_1$. In a Bloch sphere, the upper qubit rotates around the x-axis by an angle proportional to $W_1$ [4, 5].

These are simply regular Rabi oscillations for a single qubit. However, we will observe a difference if we change the state of the lower qubit to |1>, i.e., $\varepsilon_L \gg J \gg \Delta_L > 0$, while keeping the rest of the system unchanged. As shown in Fig. 1(c), when the lower qubit is in the |1> state, the upper qubit's balance point (indicated by balance line between |01> and |11>) shifts toward higher energies by an amount $J$ compared with the case in which the lower qubit is in the |0> state (balance line between |00> and |10>). As a result, the pulse amplitude that drives the upper qubit exactly to its balance point when the lower qubit is in state |0> can no longer drive the qubit to its balance point when the lower qubit is in state |1>. We thus expect the Rabi oscillations to disappear [25].

The experiment clearly demonstrates the above effect. In Figs. 1(e) and (f), we present the Rabi oscillations of the upper qubit conditional on the lower qubit's state. The x-axis corresponds to the pulse width, $W_1$. The y-axis corresponds to the lower qubit detuning, $\varepsilon_L$. Fig. 1(e) presents the differential current of the upper qubit, and Fig. 1(f) presents that of the lower qubit. Fig. 1(f) reveals that the lower qubit switches between states as the line $V_{L4} \approx -0.525$ V is crossed. When $V_{L4} \ll -0.525$ V, the lower qubit is in the |0> state, and Fig. 1(e) presents the Rabi oscillations of the upper qubit with a frequency equal to $\Delta_{U} =$ 6.2 GHz. When $V_{L4} \gg -0.525$ V, the lower qubit switches to the |1> state, and in Fig. 1(e), it is evident that the Rabi

oscillations of the upper qubit disappear. Near the balance point, i.e., when –0.526 V < $V_{L4}$ < –0.524 V, the two qubits should rotate as an entangled state, exhibiting Rabi oscillations at two frequencies. However, the two compound frequencies are outside of the range that we can detect. In any case, this entanglement is irrelevant to our CNOT gate and will be addressed elsewhere.

Figs. 1(e) and (f) demonstrate that we can completely suppress the upper qubit's Rabi oscillations by switching the lower qubit from the |0> state to the |1> state. A CNOT gate, the logical operation of which is to flip the upper qubit if the lower qubit is in state |0> and to do nothing if the lower qubit is in state |1>, can thus obtain its maximum fidelity. We perform theoretical simulations by numerically solving the master equations. Details are provided in the supplemental materials. The simulation successfully reproduces phenomena such as those observed in Figs. 1(e) and (f) when the experimentally obtained parameters were used, including $J$ = 119 μeV [26].

If we reduce $J$, for instance to 25 μeV, which is the same magnitude as $\Delta_U$ and $\Delta_L$, our simulation indicates that in this case, the upper qubit's Rabi oscillations cannot be completely suppressed. There are leakage Rabi oscillations with a 55% amplitude when the lower qubit is switched from state |0> to state |1>. Therefore, a CNOT gate for $J$ = 25 μeV will achieve a fidelity of no more than 1 - 55% = 45%. These leakage Rabi oscillations at low J occur because the two balance lines have finite line widths, as shown in Fig. 1 (c). If $J$ is smaller than or comparable to this line width, the two balance lines are smeared out, and the same voltage pulse can drive the upper qubit to its balance point regardless of whether the lower qubit is in state |0> or state |1>. Only if $J$ is much larger than this line-width will there be no overlap between these two balance lines and thus no leakage Rabi oscillations. The $J$ value required to completely separate the two balance lines is therefore the threshold value for the CNOT gate to achieve maximum fidelity.

We can calculate the dependence of the upper bound of the CNOT gate fidelity on the inter-qubit coupling strength $J$ through simulations. Details are provided in the supplemental materials. We present the process-independent fidelity without the dephasing effect in Fig. 1(g). Two important features are apparent: the fidelity increases with increasing $J$ and eventually saturates. In our case, $J$ = 119 μeV, we should, in principle, achieve 97% fidelity for the CNOT gate. However, the estimation presented in Fig. 1(g) is excessively idealistic. It assumes an infinitely long dephasing time, 100% fidelity for the single-qubit gates, and perfect pulse shaping for the two-qubit gates. As we will see, in our experiment, we achieve 68% fidelity for a CNOT gate. However, Fig. 1(g) strongly indicates that $J$ is not the major limiting factor in preventing the achievement of perfect fidelity in our experiment. The inter-qubit coupling strength in our device has already been necessarily large to achieve a satisfactory CNOT gate. This is the greatest advancement of this study with respect to earlier experiments.

**2 Pulse Shaping**

Additional experimental challenges remain in the development of a functional CNOT gate. The voltage pulses required for the implementation of a single-charge-qubit gate are already very short (200 – 500 ps). To demonstrate a CNOT gate, we require up to three sequential ultra-short pulses. These pulses must be carefully synchronized and aligned. Here, we demonstrate how we manipulate two pulses, one on the lower qubit and the other on the upper qubit, to coherently rotate the lower qubit and thus control the state of the rotation of the upper qubit. Further details regarding pulse shaping are presented in the supplemental materials.

A schematic description of the manipulation process is presented in Fig. 2(a). Both qubits are initialized in state |0>. In addition to a rectangular pulse of width $W_1$ on the upper qubit, as described above, another rectangular pulse of width $W_2$ is applied to one gate of the lower qubit, $L_5$. The lower pulse ($W_2$) is applied first, and the upper pulse ($W_1$) follows after a delay time (approximately 100 ps) that is much shorter than the dephasing time (approximately 1200 ps). If the lower pulse is terminated at $2n\pi$, then the lower qubit will remain in the |0> state. The upper pulse will then rotate the upper qubit by an angle proportional to $W_1$. By contrast, if the lower pulse is terminated at $2(n+1)\pi$, then the lower qubit will enter the |1> state. Consequently, the upper pulse will have no effect, and the upper qubit will remain in the |0> state regardless of $W_1$.

Generally, we assume that the pulse of width $W_2$ rotates the lower qubit by an angle $\beta$ and that the pulse of width $W_1$ rotates the upper qubit by an angle $\alpha$ when the lower qubit is in state |0>. Then, the two qubits will end up in the following entangled state: $cos\alpha\ cos\beta$ |00> + $sin\alpha\ cos\beta$ |10> + $sin\beta$ |01>. The probability of finding the upper qubit in state |0> is $P_U^0 = 1 - sin^2\alpha\ cos^2\beta$, and the probability of finding the lower qubit in state |0> is $P_L^0 = cos^2\beta$. Therefore, we predict that $P_U^0$ should oscillate with both $W_1$ and $W_2$, whereas $P_L^0$ should oscillate only with $W_2$. Moreover, the dependence of $P_U^0$ on $W_2$ is out of phase by $\pi$ compared with $P_L^0$. We simulate this process by solving the master equations, as shown in Figs. 2(b) and (c) [16].

Experimentally, we observe the predicted pattern shown in Figs. 2(d) and (e). The QPC differential current for the lower qubit periodically oscillates only along the $W_2$ axis, whereas that for the upper qubit exhibits oscillations along both the $W_1$ and $W_2$ axes. The oscillation frequencies along the $W_1$ and $W_2$ axes are $\Delta_U$ and $\Delta_L$, respectively. In addition, the dependence on $W_2$ is out of phase by approximately $\pi$ between the upper and lower qubits, which is as predicted. This finding demonstrates that we can indeed coherently control the Rabi oscillations of the upper qubit.

In addition, the QPC differential current for the lower qubit is invariant with respect to the upper pulse of width $W_1$. This observation serves as a proof that there is no observable crosstalk between the two qubits.

## 3 CNOT Processing Tomography

Based on the achievement of sufficiently high $J$ and proper pulse shaping, we now test the operation of a CNOT gate and perform tomography measurements to determine its processing fidelity [8, 14-19]. Figure 3(a) presents the process flowchart. In the initialization process, we reset the two qubits to the |00> state. Then, in the input preparation process, we apply certain pulses to both the upper and lower qubits to obtain different input states. By tuning the pulse widths $W_2$ and $W_3$, we prepare four input states: |00>, |10>, |01>, and |11>. Finally, these input states are fed into the CNOT gate, which consists of a π pulse on the upper qubit. The logic of the CNOT gate means that the upper qubit undergoes a π rotation if the lower qubit is in the |0> state and no rotation if the lower qubit is in the |1> state. Therefore, after passing through the CNOT gate, the four input states will be transformed into the |10>, |00>, |01>, and |11> states, respectively.

For the first input state |00>, the initial state is directly sent to the CNOT gate without any preparatory pulse. To prepare a |10> input state, a π pulse is applied to the upper qubit prior to the CNOT gate. A π pulse on the lower qubit will yield the |01> input state. Finally, we apply a π pulse to the lower qubit followed by a π pulse with an elevated amplitude on the upper qubit to obtain a |11> input state. The purpose of elevating the upper pulse amplitude is to force the upper qubit to rotate after the lower qubit has already been switched into the |1> state. Experimentally, because π pulses are too short and inevitably cause adiabatic phenomena, we use 3π pulses instead (360 ps for the upper qubit and 390 ps for the lower qubit in our experiment).

The output of the CNOT gate for each of the four input states is read through the QPC current. We use the pulse-modulation technique developed in previous studies to convert the QPC current into a state probability [8]. Further details are provided in the supplemental materials. In Figs. 3(b) and (c), we sweep the pulse width of the CNOT gate ($W_I$) and measure the probabilities $P_U^0$ and $P_L^0$ after generating each of the four input states. As expected, $P_U^0$ exhibits Rabi oscillations for input |00>. For input |10>, the Rabi oscillations are shifted by a phase of π. For inputs |01> and |11>, $P_U^0$ exhibits essentially no oscillation because the lower qubit has been switched into state |1>. The difference between the two inputs is that $P_U^0$ remains at a high level for input |01> and at a low level for input |11>. $P_L^0$ exhibits essentially no dependence on $W_I$ because the upper pulse does not affect the lower qubit.

From Figs. 3(b) and (c), we extract the values of $P_U^0$ and $P_L^0$ at $W_I$ = 360 ps, which corresponds to a 3π pulse on the upper qubit and therefore a CNOT gate. Based on these values, we obtain the density matrix for the CNOT processing tomography, as illustrated in Fig. 3(d). For comparison, we simulated the density matrices for an ideal CNOT gate and for a CNOT gate with an inhomogeneous dephasing time of 1200 ps, as shown in Figs. 3(e) and (f), respectively. The detailed values of these density matrixes are provided below:

$$D_{measured} = \begin{pmatrix} 0.09 & 0.89 & 0.002 & 0.02 \\ 0.87 & 0.12 & 0.01 & 0.002 \\ 0.06 & 0.02 & 0.74 & 0.18 \\ 0.02 & 0.07 & 0.23 & 0.68 \end{pmatrix},$$

$$D_{predicted}^{ideal} = \begin{pmatrix} 0 & 1 & 0 & 0 \\ 1 & 0 & 0 & 0 \\ 0 & 0 & 1 & 0 \\ 0 & 0 & 0 & 1 \end{pmatrix}, \quad D_{predicted}^{T_2^* = 1200 ps} = \begin{pmatrix} 0.05 & 0.95 & 0 & 0 \\ 0.90 & 0.10 & 0 & 0 \\ 0.003 & 0.05 & 0.94 & 0.007 \\ 0.005 & 0.05 & 0.06 & 0.89 \end{pmatrix}$$

The measured fidelity is 0.68. We must reiterate that the inter-qubit coupling strength *J* is not the limiting factor responsible for this imperfect fidelity because *J* is already sufficiently large to allow us to completely switch the rotations of the upper qubit on and off by manipulating the state of the lower qubit. We believe that there are two main factors that account for the deviation of the measured fidelity from 1. First, the relatively short dephasing time of a single charge qubit causes errors in single-qubit operations, and these errors are carried over into the two-qubit operations. Once the lower qubit suffers dephasing, the lower qubit state will contain a |0> component even when it should be in the |1> state. This leakage to the |0> state will cause the suppression of the upper qubit's Rabi oscillations to be incomplete. In combination with the dephasing of the upper qubit, this effect will degrade the ultimate CNOT gate fidelity. By comparing Figs. 3(f) and (e), our simulation reveals that the predicted CNOT gate fidelity decreases to 89% when a qubit dephasing time of 1200 ps [5] is considered.

Second, the pulse shaping of multiple ultra-short pulses is extremely challenging and cannot be made ideal. The relatively short dephasing time in our system requires us to complete gate operations as quickly as possible. Although this forces us to boost the gate operation speed, which proves to be helpful, it also increases the risk of poor pulse shaping. In particular, we observe that the finite nature of the pulse rising and falling times makes the precise tuning of the pulse width difficult. In addition, we observe that the increase in the pulse amplitude with increasing pulse width makes it difficult to accurately control the amplitude of each pulse and to align the amplitudes of multiple pulses. The accumulation of all these errors gives rise to deviation between the final output qubit state and the desired qubit state. This might be the primary reason for the CNOT processing fidelity to drop to approximately 68%.

Nonetheless, our gate fidelity is already comparable to the maximum fidelity achieved for two electron spin qubits [22]. The relatively short dephasing time of charge qubits has always been an obstacle preventing the serious consideration of the possibility of multiple-charge qubits. However, the large intra-qubit coupling and inter-qubit coupling originating from the direct Coulomb interaction between electron charges enable us to operate the CNOT gate at a very high clock speed (a few GHz) and to maintain the gate fidelity at a satisfactory level. Coulomb interactions are significant in various types of multi-qubit systems. For example, spin qubits in silicon-based quantum dots have recently demonstrated remarkably long coherence times [27-30]. High-quality single-electron spin gates have been demonstrated using Si. However, long-range inter-qubit coupling still originates from the Coulomb interaction. Two-qubit gates using Si still suffer from high charge noise and therefore still require further development. We hope that our demonstration of two-qubit gates based on the Coulomb interaction may offer inspiration for the investigation of semiconductor multi-qubits.

There is still room to improve the fidelity of our electron charge two-qubit gates. The errors originating from imperfect pulse shaping are deterministic and could be corrected with further progress in high-frequency technology. We hope that the advancement of picosecond pulse generators and the incorporation of on-chip transmission lines will help us to improve the fidelity of our single- and double-qubit gates. Moreover, the qubit dephasing effect is intrinsic, and new materials or architectures will be necessary to achieve significant improvements. Recent progress in the field of hybrid qubits has demonstrated that fast operating speeds and long coherence times can be simultaneously achieved in electron charge qubits by engineering an energy structure with certain excited states [8]. If similar schemes can be applied to increase $T_2^*$ in our system by perhaps 1 to 2 orders of magnitude, the fidelity of the two-qubit CNOT gate could dramatically increase.

**4 Building Blocks of Universal Two-Qubit Gates**

In the CNOT tomography measurement, we considered only the amplitudes of the quantum states. However, Fig. 2 previously illustrated the quantum nature of the CNOT gate, which has no counterpart among classical gates. The amplitudes of the quantum states of both the upper and lower qubits can be set to any arbitrary superposition value, corresponding to rotations by arbitrary angles around the x-axis in each Bloch sphere. During this process, the entanglement of the two-qubit states is established, and the CNOT logic holds for these quantum states.

Furthermore, we will show that we can vary both the phase and amplitude of the qubit's states while preserving the CNOT logic. As illustrated in Fig. 4(a), a pulse with a fixed width of 100 ps is applied to the lower qubit. This pulse width is shorter than the rise and fall times combined, and therefore, the pulses can be regarded as triangular. We initialize the lower qubit in state |0> and sweep the lower pulse amplitude. Alternatively, we can fix the pulse amplitude and sweep the lower qubit's detuning. In both cases, the lower qubit is driven to pass through its balance point if the pulse amplitude is larger than its detuning. This adiabatic passage through the balance point induces the Landau-Zener-Stuckelberg (LZS) effect, corresponding to rotation around both the x- and z-axis in the Bloch sphere [6].

Immediately following the lower pulse, a rectangular pulse is applied to the upper qubit and induces Rabi oscillations in the upper qubit. Here, we demonstrate that the Rabi oscillations of the upper qubit are controlled by the lower qubit's phase accumulation caused by LZS interference. First, let us suppose that the triangular pulse can independently drive the lower qubit from the |0> state into the $U(\beta,\psi)$ |0> + $V(\beta,\psi)$ |1> state, where $U^2(\beta,\psi) + V^2(\beta,\psi) = 1$, and that the rectangular pulse can independently drive the upper qubit from the |0> state into the $cos\alpha$ |0> + $sin\alpha$ |1> state if the two qubits are completely uncorrelated.

In reality, the two qubits are coupled, and the CNOT gate logic ensures that the upper triangular pulse can only cause the upper qubit to rotate when the lower qubit is in the |0> state. Consequently, after both the lower and upper pulses, the final entangled two-qubit state will be as follows: $U(\beta,\psi) cos\alpha$ |00> + $U(\beta,\psi) sin\alpha$ |10> + $V(\beta,\psi)$ |01>. The probability of finding the upper qubit in state |0> is $P_U^0 = 1 - U^2(\beta,\psi) sin^2\alpha$, and the probability of finding the lower qubit in state |0> is $P_L^0 = U^2(\beta,\psi)$. $U^2(\beta,\psi)$ oscillates with both the amplitude (through $\beta$) and phase (through $\psi$) of the lower qubit's state. However, the oscillation of the phase is much faster than that of the amplitude. Therefore, in the time window of our experiment, we predominantly observe periodic oscillations with the phase. Thus, $P_U^0$ will exhibit cosine oscillations with both $W_1$ (through $\alpha$) and $A_2$ or $E_2$ (mainly through $\psi$), and $P_L^0$ will oscillate only with $A_2$ or $E_2$ (predominantly through $\psi$). Again, we note that the dependence of $P_U^0$ on $A_2$ or $E_2$ is out of phase by $\pi$ compared with that of $P_L^0$. In Figs. 4(b) and (c), we present the simulated responses of $P_U^0$ to $A_2$ and $E_2$, respectively.

This interpretation explains the data depicted in Figs. 4(d) and (e), where we present the oscillations of the upper qubit with $W_1$ and $A_2$ or $E_2$. As expected, the upper qubit exhibits not only Rabi oscillations with respect to $W_1$ through angle $\alpha$ but also oscillations with $A_2$ or $E_2$ through the phase $\psi$. The oscillation of the upper qubit with the phase of the lower qubit indicates that the phase of the lower qubit's state can be used to control the state of the upper qubit. Our CNOT gate is thus proven to operate on quantum states of the qubits. In the supplemental materials, we demonstrate that we can even rotate the phase and amplitude of both the upper and lower qubits while preserving the CNOT gate quantum logic.

More importantly, arbitrary quantum logic gates can, in principle, be implemented using a combination of a set of universal single-qubit gates and any one two-qubit gate, such as a CNOT gate. In our system, we can achieve universal single-qubit gates for both qubits using the LZS effect [6], and we can also achieve a CNOT two-qubit gate. It should be able to behave as a universal two-qubit gate, and in principal, it should be possible to use this gate to construct universal multi-qubit gates. The key advantage of our system is that the gate operations for single, double, and multiple qubits are all based on the same type of coupling: the Coulomb interaction among electron charges.

**Summary**

In summary, a string of technical accomplishments, including the achievement of a strong inter-qubit coupling and the synchronization of multiple ultrafast-pulses, enabled us to demonstrate universal two-qubit operations in an all-electrically-controlled semiconductor charge system. CNOT processing tomography shows high gate processing and read-out fidelities comparable to that of electron spin two-qubit gates. Our finding appears to be contrary to the conventional perception of that charge qubits are inferior compared to spin qubits in semiconducting materials. At current stage, trading shorter dephasing time for faster qubit operation time, charge qubits can perform equivalently well in the two-qubit level. Optimistically, we argue that with a better control of pulse-shape and a better design the dispersion relations, which are completely deterministic, the semiconductor charge qubits may become a force to contend with in the scalable quantum computation arena.

**Methods**

The two-DQD device was defined via electron-beam lithography on a molecular-beam-epitaxially grown GaAs/AlGaAs heterostructure. A two-dimensional electron gas (2DEG) is present 95 nm below the surface. The 2DEG has a density of $3.2 \times 10^{11}$ cm$^{-2}$ and a mobility of $1.5 \times 10^5$ cm$^2$/Vs. Figure 1a presents a scanning electron micrograph of the surface gates. Five upper gates $U_1$-$U_5$ and two horizontal gates $H_1$ and $H_2$ form the upper DQD. Five lower gates $L_1$-$L_5$ and two horizontal gates $H_1$ and $H_2$ form the lower DQD. The horizontal gates $H_1$ and $H_2$ also tune the capacitive coupling strength between the upper and lower DQDs. Direct electron tunneling is suppressed between the two DQDs by ensuring adequate negative bias voltages on gates $H_1$ and $H_2$. The four gates $Q_1$-$Q_4$ define QPCs for the monitoring of the charge status on each DQD. The experiments are performed in an Oxford Triton dilution refrigerator with a base temperature of 10 mK. Two Agilent 81134A pulse generators, which have a rise time of 65 ps and a time resolution of 1 ps, are used to deliver fast pulse trains through semi-rigid coaxial transmission lines to the device. Standard lock-in modulation and detection techniques are used for the charge-sensing readout. Through electronic transport and photon-assisted-tunneling (PAT) measurements, wherein the electron energy can be read directly from the source-drain bias voltage and the photon frequency, we conclude that the energy-voltage lever arm is approximately 100 μeV/mV for the barrier gates ($U_1$, $U_5$, $L_1$, and $L_5$) and 30 μeV/mV for the plunger gates ($U_2$, $U_4$, $L_2$, and $L_4$).

**References**

1. M. A. Nielsen, and I. L. Chuang. Quantum computation and quantum information. (Cambridge Univ. Press, 2000).
2. R. Hanson, et al. Spins in few-electron quantum dots. Rev. Mod. Phys. 79, 1217 (2007).
3. F. A. Zwanenburg, et al., Silicon quantum electronics. Rev. Mod. Phys. 85, 961 (2013).
4. T. Hayashi, et al. Coherent manipulation of electronic states in a double quantum dot. Phys. Rev. Lett. 91, 226804 (2003).
5. K. D. Petersson, et al. Quantum coherence in a one-electron semiconductor charge qubit. Phys. Rev. Lett. 105, 246804 (2010).
6. G. Cao, et al. Ultrafast universal quantum control of a quantum-dot charge qubit using Landau-Zener-Stuckelberg interference. Nature Commun. 4, 1401 (2013).
7. Z. Shi, et al. Fast coherent manipulation of three-electron states in a double quantum dot. Nat. Commun. 5, 3020 (2014).
8. Dohun Kim, et al. Quantum control and process tomography of a semiconductor quantum dot hybrid qubit, Nature 511, 70 (2014).
9. Dohun Kim, et al. Microwave-driven coherent operation of a semiconductor quantum dot charge qubit. arXiv:1407.7607 (2014)
10. K. D. Petersson, et al. Microwave-driven transitions in two coupled semiconductor charge qubits. Pyhs. Rev. Lett. **103**, 016805 (2009).
11. Gou Shinkai, et al. Correlated coherent oscillations in coupled semiconductor charge qubits. Phys. Rev. Lett. 103, 056802 (2009).
12. T. Fujisawa, et al. Multiple two-qubit operations for a coupled semiconductor charge qubit. Physica E 43 730-734 (2011).
13. I. van Weperen, et al. Charge-state conditional operation of a spin qubit. Pyhs. Rev. Lett. 107, 030506 (2011).
14. Yu. A. Pashkin, et al. Quantum oscillations in two coupled charge qubits. Nature 421, 823 (2003).
15. T. Yamamoto, et al. Demonstration of conditional gate operation using superconducting charge qubits. Nature 425, 941 (2003).
16. Mika A. Sillanpaa, et al. Coherent quantum state storage and transfer between two phase qubits via a resonant cavity. Nature 449, 438 (2007).
17. L. DiCarlo, et al. Demonstration of two-qubit algorithms with a superconducting quantum processor. Nature 460, 240 (2009).
18. R. C. Bialczak, et al. Quantum process tomography of a universal entangling gate implemented with Josephson Phase Qubits. Nature Physics 6, 409 (2010).
19. J. H. Plantenberg, et al. Demonstration of controlled-NOT quantum gates on a pair of superconducting quantum bits. Nature 447, 836 (2007).
20. R. Brunner, et al. Two-qubit gate of combined single-spin rotation and interdot spin exchange in a double quantum dot. Pyhs. Rev. Lett. 107, 146801 (2011).
21. K. C. Nowack, et al. Single-shot correlations and two-qubit gate of solid-state spins. Science 333, 1269 (2011).
22. M. D. Shulman, et al. Demonstration of entanglement of electrostatically coupled

singlet-triplet qubits. Science 336, 202 (2012).

23. Normally, the eigenstates are different from the charge states (|RR>, |LR>, |LR> and |LL>) that the QPCs can detect, except when the detuning of each qubit is far from its balance point ($|\varepsilon_{U,L}| \gg 0$). However, in our gate operations, we always initialize the states far from the balance points, and we allow the initialized states to adiabatically evolve into the eigenstates before gate operations and then to adiabatically evolve back into the charge states once gate operations are terminated. Therefore, in subsequent discussions, we will ignore the unitary transformation between the eigenstates and the charge states and will simply discuss the gate operations in the basis of the eigenstates.

24. The pulse's combined rise and fall time is measured to be 130 ps on top of the refrigerator. When we fix the pulse width shorter than or close to 130 ps and sweep the pulse amplitude or energy detuning, Landau-Zener-Stuckelberg interference originating from adiabatic evolution is seen. When we sweep the pulse width to longer values, non-adiabatic evolution such as Rabi oscillations is observed.

25. For simplicity, in this paper, we present the details only for the control of the upper qubit through the manipulations of the lower qubit. Considering the symmetric design, the opposite would certainly be possible, i.e., controlling the lower qubit by manipulating the upper qubit.

26. Sometimes, we present the QPC differential current because of its better signal-to-noise ratio. Our simulation focuses on the state probability, which is directly related to the QPC current. However, the simulation reflects the same features as those of the output signal. Moreover, we also measure the state probability directly, as discussed in the section concerning the CNOT processing tomography.

27. B. M. Maune, et al. Coherent singlet-triplet oscillations in a silicon-based double quantum dot. Nature 481, 344 (2012).

28. Jarryd J. Pla, et al. A single-atom electron spin qubit in silicon. Nature 489, 541 (2012).

29. Jarryd J. Pla, et al. High-fidelity readout and control of a nuclear spin qubit in silicon. Nature 496, 334 (2013).

30. E. Kawakami, et al. Electrical control of a long-lived spin qubit in a Si/SiGe quantum dot. Nature Nanotechnology 9, 666 (2014).

**Acknowledgments**

This work was supported by the National Fundamental Research Program (Grant No. 2011CBA00200), the National Natural Science Foundation (Grant Nos. 11222438, 11174267, 61306150, 11304301, 11274294, and 91121014), and the Chinese Academy of Sciences.

**Author contributions**

H. O. Li and G. P. Guo performed the measurements. G. Cao, G. C. Guo, and H. W. Jiang provided theoretical support and analyzed the data. G. D. Yu and M. Xiao fabricated the samples. M. Xiao and G. P. Guo supervised the project. All authors contributed to write the paper.

**Figure 1 Strong Inter-qubit Coupling** (a) Scanning electron micrograph of the device. (b) There exists an exchange energy *J* originating from Coulomb repulsion. (c) and (d) Experimentally measured *J*. (e) and (f) The manipulation of the state of the lower qubit can completely switch the Rabi oscillations of the upper qubit on and off. (g) Process-independent CNOT gate fidelity as a function of *J*, calculated by solving the master equations.

**Figure 2 Pulse Shaping** (a) Gate pulse flowchart for the coherent control of the upper qubit's Rabi oscillations through the pulse driving of the lower qubit. (b) and (c) Theoretical simulations of $P_U^0$ and $P_L^0$, respectively. The red dashed lines indicate two adjacent valleys of the upper qubit's oscillations with respect to $W_1$. The yellow dashed lines indicate two adjacent valleys of the oscillations with respect to $W_2$ for $P_U^0$, whereas indicate two adjacent peaks for $P_L^0$. (d) and (e) Experimentally observed differential current of the upper and lower QPCs, respectively.

**Figure 3 CNOT Processing Tomography** (a) Gate pulse flowchart for the CNOT processing tomography measurements. (b) and (c) Probabilities for the upper and lower qubits, respectively, as functions of the pulse width $W_1$. For the actual CNOT gate, $W_1$ = 360 ps. The red circles, pink stars, blue triangles, and green crosses correspond to input states |00>, |10>, |01>, and |11>, respectively. (d) Experimentally measured probabilities for the CNOT output states. The red, pink, blue, and green bars correspond to input states |00>, |10>, |01>, and |11>, respectively. (e) and (f) Theoretically predicted probabilities: (e) corresponds to no dephasing and (f) corresponds to a dephasing time of 1200 ps.

**Figure 4 Building Blocks of Universal Two-qubit Gates** (a) Gate pulse flowchart for the use of the lower qubit's phase to control the upper qubit's Rabi oscillations. (b) and (c) Theoretical simulations of $P_U^0$ in which we sweep the lower qubit's pulse amplitude and detuning, respectively, to control the Rabi rotations of the upper qubit. The red dashed lines indicate two adjacent valleys of the oscillations of the upper qubit with respect to $W_1$. The yellow dashed lines indicate two adjacent valleys in the response of the upper qubit to $A_2$ or $\varepsilon_L$. (d) and (e) Experimental results for the differential current of the upper QPC.

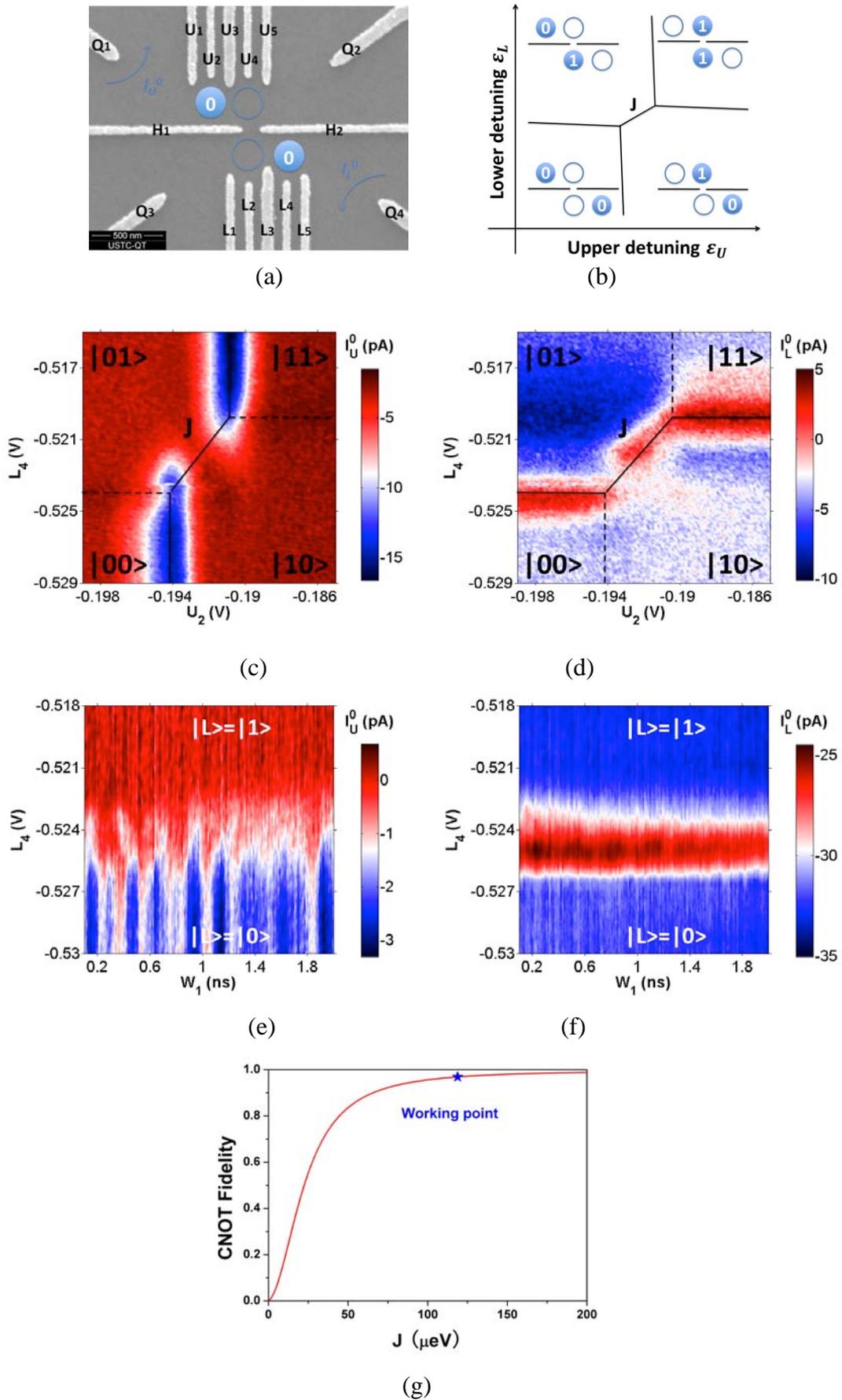

(a) (b)

(c) (d)

(e) (f)

(g)

**Figure 1 Strong Inter-qubit Coupling**

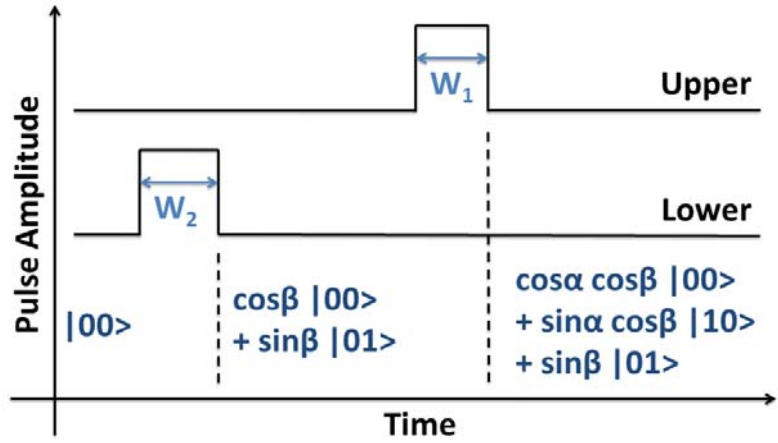

(a)

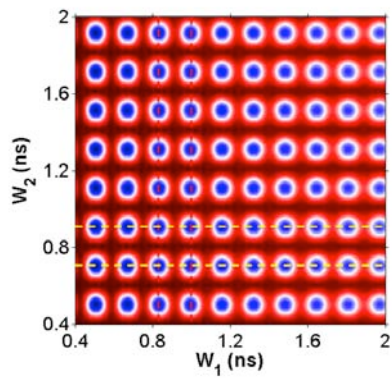

(b)

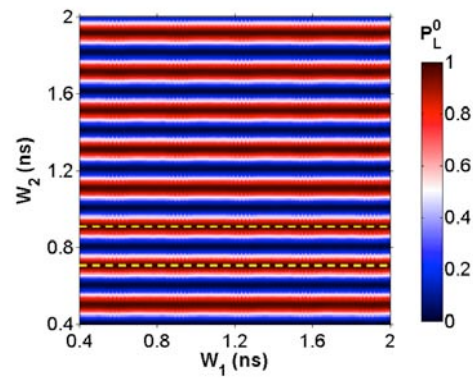

(c)

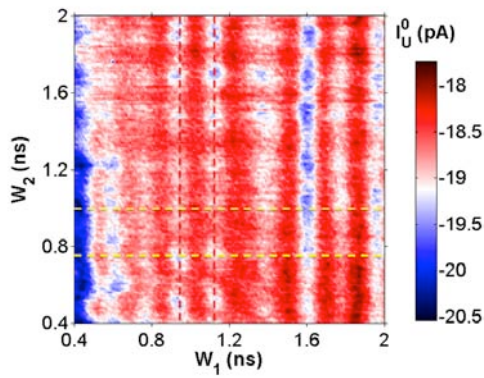

(d)

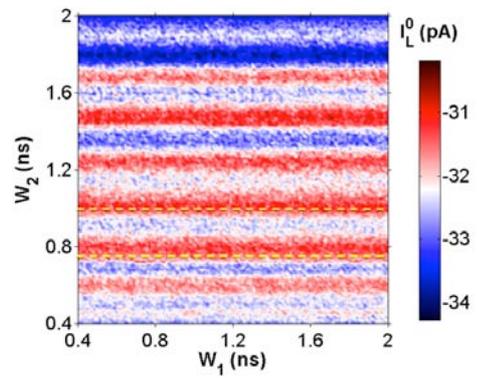

(e)

**Figure 2 Pulse Shaping**

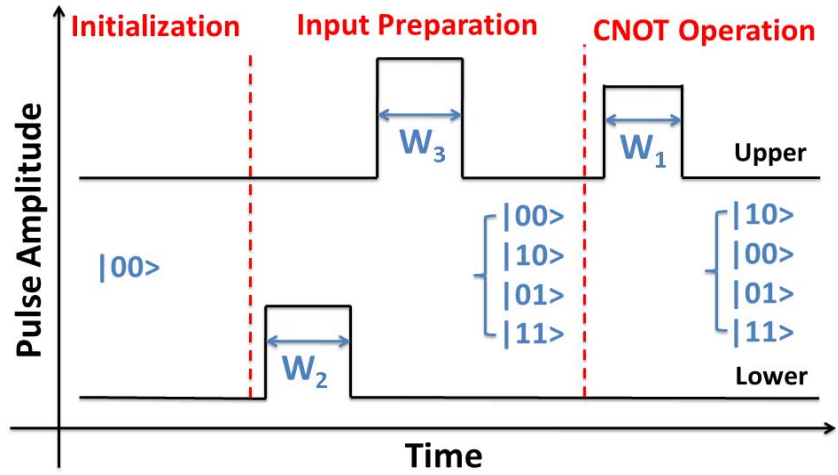

(a)

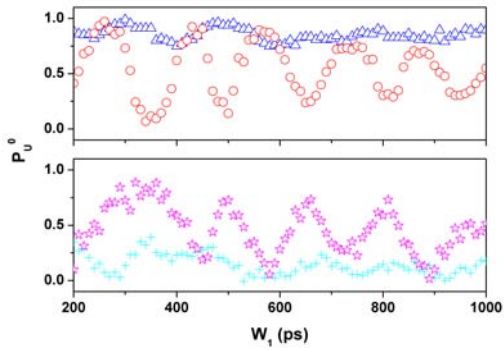

(b)

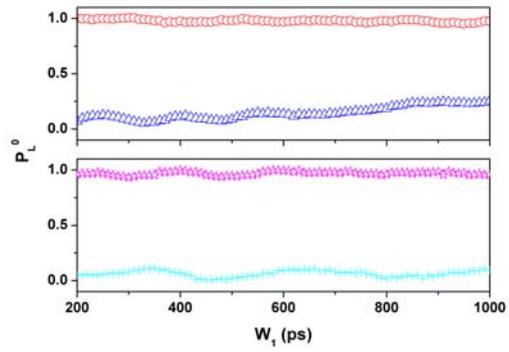

(c)

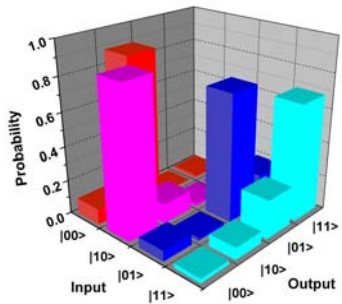

(d)

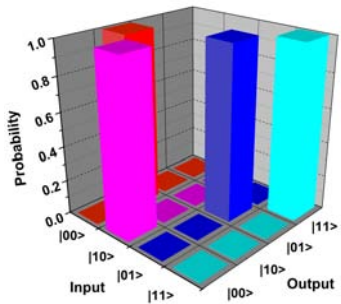

(e)

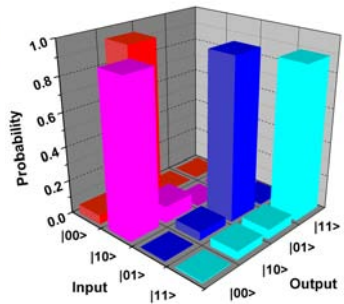

(f)

**Figure 3 CNOT Processing Tomography**

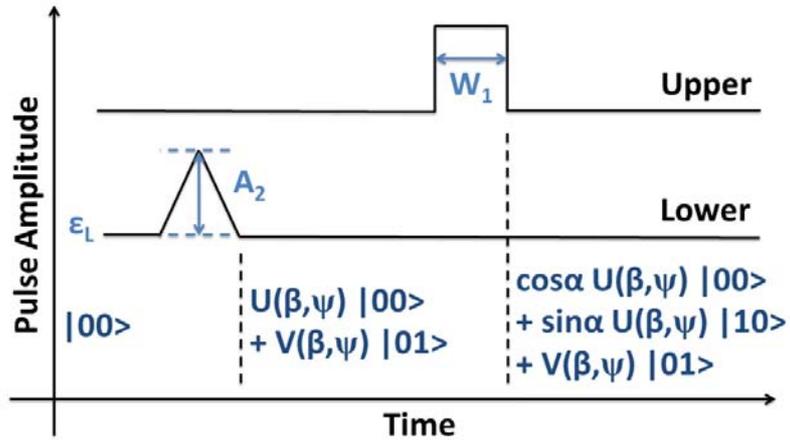

(a)

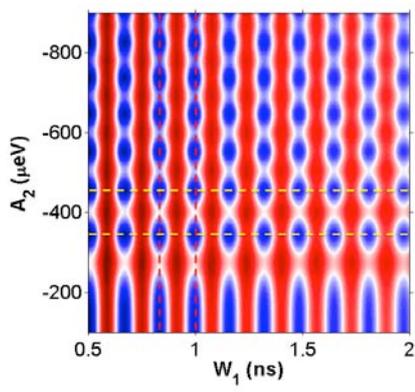

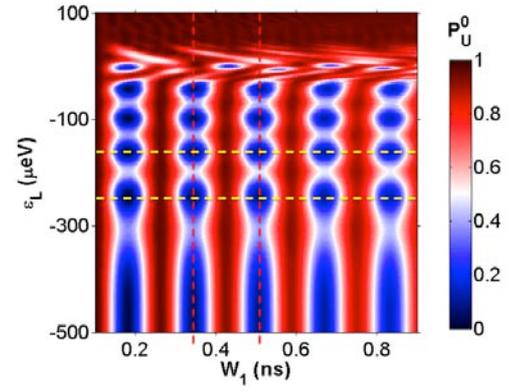

(b)                (c)

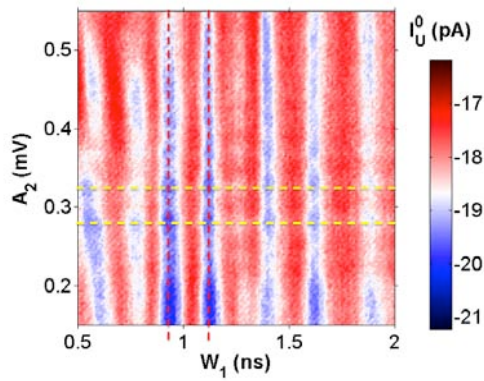

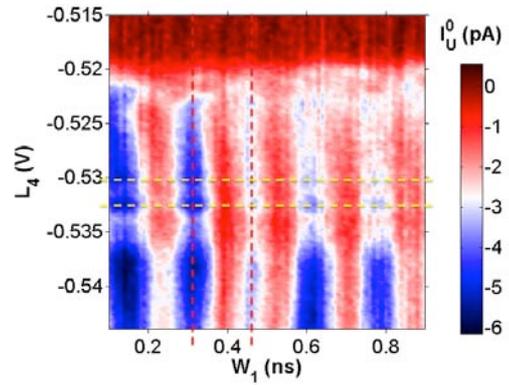

(d)                (e)

**Figure 4 Building Blocks of Universal Two-qubit Gates**

# Supplemental Materials:

# Controlled-Not Quantum Logic Gate in Two Strongly Coupled Semiconductor Charge Qubits

## S1 Single-qubit Manipulation and QPC Measurement

We independently consider the charge energy diagrams for the upper and lower DQDs, as shown in Figs. S1(a) and (b), respectively. The electron charge occupation on either side of the upper DQD is tuned by two upper plunger gates, $U_2$ and $U_4$, and detected by the upper QPC. In the same way, the charge occupation on either side of the lower DQD is tuned by two lower plunger gates, $L_2$ and $L_4$, and detected by the lower QPC.

We operate the two-qubit system between charge occupations of (2,1) and (1,2) for the upper DQD and between charge occupations of (1,0) and (0,1) for the lower DQD, as indicated by the circles in Figs. S1(a) and (b). The upper energy detuning $\varepsilon_U$, obtained through the voltage detuning between $U_2$ and $U_4$, controls the (2,1) ←→ (1,2) transition for the upper DQD. Similarly, the lower energy detuning $\varepsilon_L$, obtained through the voltage detuning between $L_2$ and $L_4$, controls the (1,0) ←→ (0,1) transition for the lower DQD.

Generally, we measure the differential current of the QPC because of its better signal-to-noise ratio. In two-qubit gate operations, we need to know the state probability explicitly to measure the fidelity. For this purpose, we modify the QPC measurement technique [S1]. The pulse voltage on gate $U_1$, which is used to induce coherent oscillations, is modulated by a 200 Hz square wave. The same 200 HZ square wave also triggers the lock-in that is used to measure the modulated QPC current. Therefore, the ratio between the lock-in-measured oscillation amplitude and the largest oscillation amplitude tells us the state probability if we regard the largest oscillation amplitude as having a probability 1.

In Fig. S1(c), the blue dotted line represents a coherent oscillation curve of the upper qubit. Here, the upper detuning is tuned such that the upper pulse drives it exactly to its balance point. We can see that there is an overall shift in background with pulse width. Different pulse widths result in different average voltages on gate $U_1$. Because the QPC is operated approximately within its most sensitive range, which is not strictly linear, a shift in $U_1$ results in a shift in the sensitivity of the QPC current. This is most likely the origin of the background shift with respect to the pulse width. We can eliminate this shifting background by dividing the QPC signal by a pure background, which can be obtained at lower detuning value where Rabi oscillations do not occur.

By dividing the raw Rabi oscillation by the background, we can obtain the normalized state probability for Rabi oscillation, as indicated by the black dotted curve in Fig. S1(d). This curve can be well fitted with a decaying cosine oscillation: $a_0 \exp(-(W_1/T_2^*)^2) \cos(2\pi W_1/\Delta_U + b_0) + a_1 W_1 + a_2$. Through fitting, we obtain $T_2^* = 1200$ ps, $\Delta_U = 6.2$ GHz, $a_0 = 0.50$, $a_1 = 0$, $a_2 = 0.50$, and $b_2 = 0.03\pi$.

## S2 Simulation through Solving the Master Equations

The Hamiltonian described in equation (1) can be written in matrix form as follows:

$$H_{2q} = \frac{1}{2} \begin{vmatrix} \varepsilon_U + \varepsilon_L & \Delta_L & \Delta_U & 0 \\ \Delta_L & \varepsilon_U - \varepsilon_L & 0 & \Delta_U \\ \Delta_U & 0 & -\varepsilon_U + \varepsilon_L & \Delta_L \\ 0 & \Delta_U & \Delta_L & -\varepsilon_U + \varepsilon_L + 2J \end{vmatrix} \quad (S1)$$

The evolution of the two-qubit system can be described by a time-dependent 4x4 density matrix $\rho$, which obeys the master equations:

$$\frac{d\rho}{dt} = -\frac{i}{\hbar}[H_{2q}, \rho] \quad (S2)$$

To numerically solve these equations, we must first separate the real and imaginary parts of the density matrix. Let us define $W = real(\rho) + imaginary(\rho)$. $W$ is a real matrix and $trace(W) = 1$. Because both $\rho$ and $H_{2q}$ are Hermitian and $H_{2q}$ is real, we can transform the master equations in terms of $\rho$ into those in terms of $W$:

$$\frac{dW}{dt} = -\frac{i}{\hbar}[H_{2q}, W^T] \quad (S3)$$

Here $W^T$ is the transpose of $W$. The transformed master equations consist of 15 correlated real differential equations and can be numerically solved using popular scientific programming languages such as MATLAB or Python. The diagonal elements of $W$ are identical to those of $\rho$ and contain all the information that we need:

$$P_U^0 = \rho_{1,1} + \rho_{3,3} = W_{1,1} + W_{3,3} \quad (S4)$$

$$P_L^0 = \rho_{1,1} + \rho_{2,2} = W_{1,1} + W_{2,2} \quad (S5)$$

The initial conditions for $W$ and $\rho$ will also be the same if we initialize $\rho$ to be real. For two-qubit operations, we always begin with conditions such that $|\varepsilon_{U,L}| \gg 0$ for the purpose of initializing the two-qubit system in the $|00\rangle$ state. Considering the thermal activation at finite temperature, the initial conditions should be as follows:

$$W(t=0) = \rho(t=0) = \frac{diagonal(e^{-E/kT})}{trace(diagonal(e^{-E/kT}))} \quad (S6)$$

Here, $E$ is a matrix whose $i^{th}$ diagonal element is the $i^{th}$ eigenvalue of $H_{2q}$. The relaxation and dephasing mechanisms can also be incorporated:

$$\frac{dW}{dt} = -\frac{i}{\hbar}[H_{2q}, W^T] - \Gamma_1(W - W(t=0)) - \Gamma_2 W \quad (S7)$$

$\Gamma_1$ is the relaxation rate matrix. $\Gamma_2$ is the dephasing rate matrix. Because we can measure the relaxation rate by varying the sampling frequency of the operating gate pulses and then numerically compensate for the relaxation effect, we ignore $\Gamma_1$ in our simulation. The dephasing rate gives rise to decaying probabilities and reduced gate fidelities. Through comparison with the experiment, we conclude that the inhomogeneous dephasing time $T_2^* = 1/\Gamma_2$ at the balance point should be approximately 1200 ps, which is of the same magnitude as the dephasing times in previous experiments [S2].

Using the initial conditions (S6), we numerically integrate the differential equations (S7) over time, thereby solving for the matrix elements of $W$ at any given time. Then, using equations (S4) and (S5), we complete the simulation of this two-qubit system. By performing parallel computation using a graphical processing unit (GPU), we can tremendously increase the speed of our simulation. Furthermore, our method can be conveniently extended to multi-qubit systems.

As an example, we demonstrate how we simulate the expected CNOT processing fidelity. First, we evaluate the CNOT fidelity in a simple manner. The essence of a CNOT gate lies in the complete on/off switching of the upper qubit's Rabi oscillations through the manipulation of the state of the lower qubit. If the dephasing issue is neglected, the Rabi oscillation amplitude of the upper qubit should decrease from 1 to 0 when the lower qubit switches from state |0> to state |1>. This behavior is observed in our experiment, as shown in Figs. 1(e) and (f). In Fig. S2(a), we simulate the case of $J$ = 119 μeV. The Rabi oscillations of the upper qubit achieve a magnitude of 0.97 when the lower qubit is in the |0> state ($\varepsilon_L \ll 0$) and drop to 0 when the lower qubit is in the |1> state ($\varepsilon_L \gg 0$).

However, the Rabi oscillations will not be completely suppressed if $J$ is not sufficiently large. This is because the balance point of the upper qubit exhibits a very small shift so that the upper gate pulse still has a chance to reach it. In Fig. S2(b), we simulate the case of $J$ = 25 μeV. There are obvious leakage Rabi oscillations when the lower qubit is in the |1> state. By varying $J$, we simulate these leakage Rabi oscillations with respect to the pulse width $W_1$. In the simulation, $\varepsilon_L$ is fixed to 200 μeV to ensure that the lower qubit is in the |1> state. Figure S2(c) clearly shows that the leakage oscillations become weaker as $J$ increases and will be suppressed when $J$ is sufficiently large.

The leakage oscillations represent the leakage probability that the upper qubit will flip even if the lower qubit is set to state |1>. Let $A_k(J)$ be this leakage probability; then, we can define the CNOT processing fidelity as $F(J) = 1 - A_k(J)$. From Fig. S2(c), for any given value of $J$, we can extract the leakage oscillation amplitude as $A_l(J)$, which is the largest leakage probability in all situations and therefore is independent of the explicit tomography process. In this manner, we can obtain $F(J)$, as shown in Fig. 1(g) and also as the red solid curve in Fig. S2(d). Evidently, $F(J)$ increases with increasing $J$ and eventually saturates.

We now consider the dephasing effect. We use $A_k'(J)$ to denote the leakage probability for the upper qubit to flip even though the lower qubit is in the |1> state in the presence of a finite qubit dephasing time. The dephasing also causes the probability for the upper (lower) pulse to flip the upper (lower) qubit, $f_U$ ($f_L$), to be less than 1. Our CNOT tomography measurement consists of four processes. We can derive the fidelity for each process: $f_U$ for input |00>, $f_U^2 + (1-f_U)^2$ for input |10>, *(1 - $A_k'$) $f_L$* for input |01>, and *(1 - $A_k'$) $f_U f_L$ + $A_k'$ (1- $f_U$) $f_L$* for input |11>. The overall CNOT gate fidelity, *F' (J),* is the minimum value among these four processes. We regard the leakage probability as $A_k'(J)$ when $W_1$ corresponds to a 3π pulse and a dephasing time of $T_2^*$ = 1200 ps is considered. The value of both $f_L$ and $f_U$ is calculated to be about 0.95 for 3π pulses when $T_2^*$ = 1200 ps. Finally, we obtain the process-dependent fidelity *F'(J),* which is presented as the green dashed curve in Fig. S2(d).

$F'(J)$ exhibits fluctuations with varying $J$, especially when $J$ is small. This is because the frequency of the leakage Rabi oscillations increases with increasing $J$. If we trace a line along the $W_l$-axis in Fig. S2(c), we will inevitably encounter fluctuations in $A_k'$ with varying $J$. Despite these fluctuations, $F'(J)$ shares the same features as $F(J)$: in general, they increase with increasing $J$ and eventually saturate.

In our experiment, $J = 119$ μeV, and we obtain $F(J) = 0.97$ and $F'(J) = 0.89$. For comparison, the black dot in Fig. S2(d) represents the experimentally observed fidelity: 0.68. This observation suggests that some other effect or effects in addition to dephasing must account for the imperfect observed CNOT fidelity. We believe that the imperfect fidelity can most likely be attributed to deficiencies in the pulse shaping of the three sequential ultra-short pulses.

**S3 Controlled-Universal-Rotations**

To rotate both the phases and amplitudes of the quantum states of the two qubits, we utilize the LZS interference effect [S3]. As illustrated in Fig. S3(a), we apply two voltage pulses, one each to the upper and lower qubits. Both pulse widths are fixed to 100 ps, shorter than the rise and fall times combined. Therefore, the pulses can be regarded as triangular. We initialize both qubits in state |0> and choose the pulse amplitude such that it will drive each qubit through its balance point with a large sweeping velocity. We can vary either the pulse amplitude or the detuning to tune the sweeping velocity. The two qubits undergo adiabatic evolutions known as the LZS effect, meaning that both qubits rotate around the x- and z-axis in each Bloch sphere. If the two qubits are completely uncorrelated, then the LZS effect should independently transform the upper qubit from the |0> state into the $U(\beta,\psi)$ |0> + $V(\beta,\psi)$ |1> state and transform the lower qubit from the |0> state into the $U(\alpha,\phi)$ |0> + $V(\alpha,\phi)$ |1> state.

Now, let us consider the coupling between the two qubits. We apply the lower pulse first. The upper pulse follows after a delay time that is much shorter than the dephasing time. Then, the x and z rotations of the upper qubit are controlled by the resulting state of the lower qubit. Consequently, two pulses transform the two-qubit state into $U(\alpha,\phi) U(\beta,\psi)$ |00> + $V(\alpha,\phi) U(\beta,\psi)$ |10> + $V(\beta,\psi)$ |01>. The probability of finding the target in the |0> state is $P_U^0 = 1 - V^2(\alpha,\phi) U^2(\beta,\psi)$, and the probability of finding the control qubit in the |0> state is $P_L^0 = U^2(\beta,\psi)$. This behavior is observed in our experiment, as shown in Figs. S3(b) and (c), where we sweep the detuning of both qubits. Theoretical simulations are presented in Figs. S3(d) and (e). The agreement between theory and experiment demonstrates that our CNOT gate functions for any quantum states of both qubits and is therefore a quantum logic gate.

This experiment demonstrates that we can rotate the phases and amplitudes of both single qubits and still achieve a CNOT gate for the combination of the two. Arbitrary quantum logic gates can, in principle, be implemented using a combination of a set of universal single-qubit gates and any one two-qubit gate such as a CNOT gate. We have demonstrated the potential of our system to serve as a universal two-qubit gate and, in principal, as a component of universal multi-qubit gates.

## S4 Pulse Synchronization

The experimental ability to control one qubit by manipulating another critically relies on the precise synchronization of multiple ultra-short voltage pulses. For a CNOT gate, at most three pulses are utilized. The two pulses applied to the upper qubit are on the same gate. Therefore, their synchronization can be performed before they are fed into the fridge using a fast oscilloscope. However, the third pulse is applied to a different qubit. A time delay between the third pulse and the other two pulses will arise as they travel through different paths to reach different gates. This system time delay remains unknown until it can be measured through its effect on the two-qubit operations.

The controlled-universal-rotations, as explained in section S3, can be utilized to synchronize the pulses on the upper and lower gates. We apply a 100-ps pulse to the control qubit and another 100-ps pulse to the target qubit. We must use ultra-short pulses for precise determination of the system time delay. The effect is that the rotation of the upper qubit is controlled by the rotation of the lower qubit, under the assumption that the upper pulse immediately follows the lower pulse. If the upper pulse comes too late after the lower pulse has finished, then qubit dephasing will cause the two qubits to behave independently.

Conversely, if the lower pulse comes immediately after the upper pulse, then the rotation of the lower qubit will be controlled by the upper qubit. If the lower pulse arrives too late compared with the upper qubit, then the rotation of both qubits will again be independent.

We modify the time delay between the pulses applied to the upper and lower qubits at the pulse generator. The unknown system time delay along the transmission path is then added to this predetermined delay to yield the final time delay. We record the coherent rotations of both qubits under these pulses, as shown in Figs. S4(a)-(f). We conclude that the system time delay is approximately +200 ps (the positive sign means that the upper pulse lags behind the lower pulse by 200 ps), because a -200 ps predetermined delay just cancels the system delay and yields independent coherent rotations for the two qubits. This number has been applied to all experiments presented in this paper.

## References

S1 Dohun Kim, et al. Quantum control and process tomography of a semiconductor quantum dot hybrid qubit, Nature 511, 70 (2014).
S2 K. D. Petersson, et al. Quantum coherence in a one-electron semiconductor charge qubit. Phys. Rev. Lett. 105, 246804 (2010).
S3 6. G. Cao, et al. Ultrafast universal quantum control of a quantum-dot charge qubit using Landau-Zener-Stuckelberg interference. Nature Commun. 4, 1401 (2013).

**Figure S1 Single-qubit Manipulation and QPC Measurement** (a) and (b) Phase diagrams for the upper and lower DQDs, respectively. (c) The blue dotted line represents the raw data of the Rabi oscillations of the upper qubit. The green solid line represents the background curve. (c) The black dotted curve represents normalized Rabi oscillations obtained by dividing the raw data by the background. The red solid line represents the theoretical fit.

**Figure S2 Simulation of CNOT Fidelity as a Function of *J*** (a) Simulated dependence of the Rabi oscillations of the upper qubit on the detuning of the lower qubit for $J = 119$ μeV. (b) Simulation for $J = 25$ eV. (c) The dependence on $J$ of the leakage Rabi oscillations of the upper qubit when the lower qubit is fixed in the |1> state ($\varepsilon_L = 200$ μeV). (d) CNOT processing fidelity as a function of $J$. The red solid curve $F$ is calculated as 1 minus the leakage oscillation amplitude in (c). The green dashed curve $F'$ is obtained for 3π pulses and $T_2^* = 1200$ ps. The black dot indicates the experimentally measured value.

**Figure S3 Controlled-Universal-Rotations** (a) Gate pulse flowchart for universal single-qubit manipulation performed in combination with the CNOT gate. (b) and (c) Experimental results for the differential currents of the upper and lower QPC, respectively. The red dashed lines indicate two adjacent valleys of the oscillations of the upper qubit with respect to $W_1$. The yellow dashed lines indicate a few adjacent valleys in response to $W_2$ for $I_U^0$, whereas indicate adjacent peaks for $I_L^0$. The red and yellow dotted lines indicate the detuning balance lines. (d) and (e) Theoretical simulations of $P_U^0$ and $P_L^0$.

**Figure S4 Pulse Synchronization** (a) and (b) Differential current of upper and lower QPCs, respectively, when the predetermined delay time from the end of pulse $W_2$ to the beginning of pulse $W_1$ is set as -100 ps at the pulse generator. The upper qubit is controlled by the lower qubit, meaning that the upper pulse lags behind the lower pulse. (c) and (d) The predetermined delay time is -200 ps. The two pulses are most likely synchronized in this case. (e) and (f) The predetermined delay time is -300 ps. The lower qubit is controlled by the upper qubit, meaning that the upper pulse is ahead of the lower pulse.

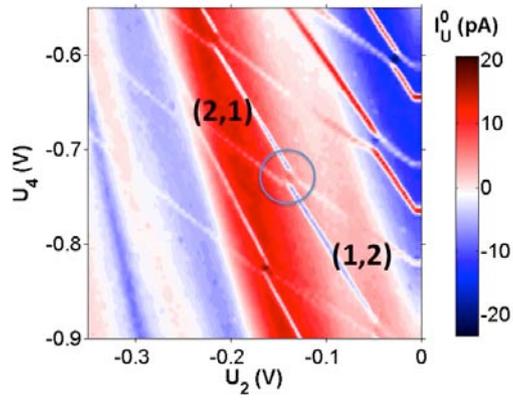 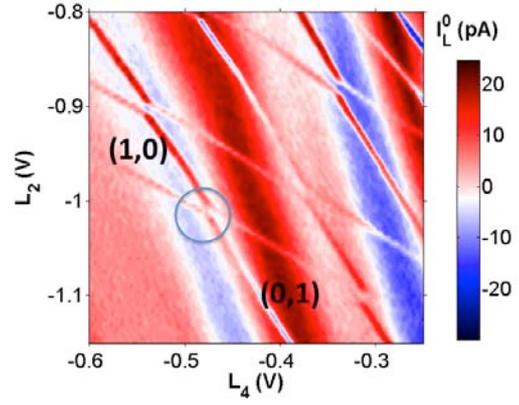

(a)                  (b)

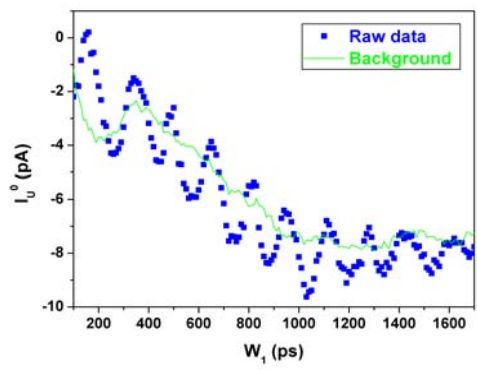 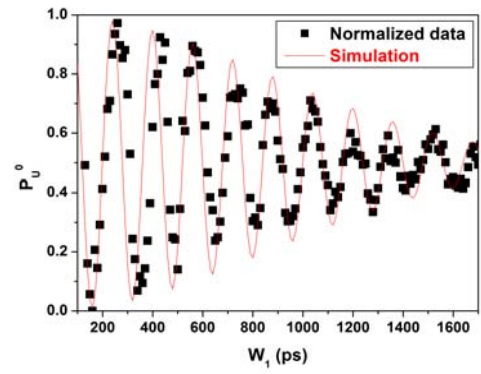

(c)                  (d)

**Figure S1 Single Qubit Manipulation and QPC Measurement**

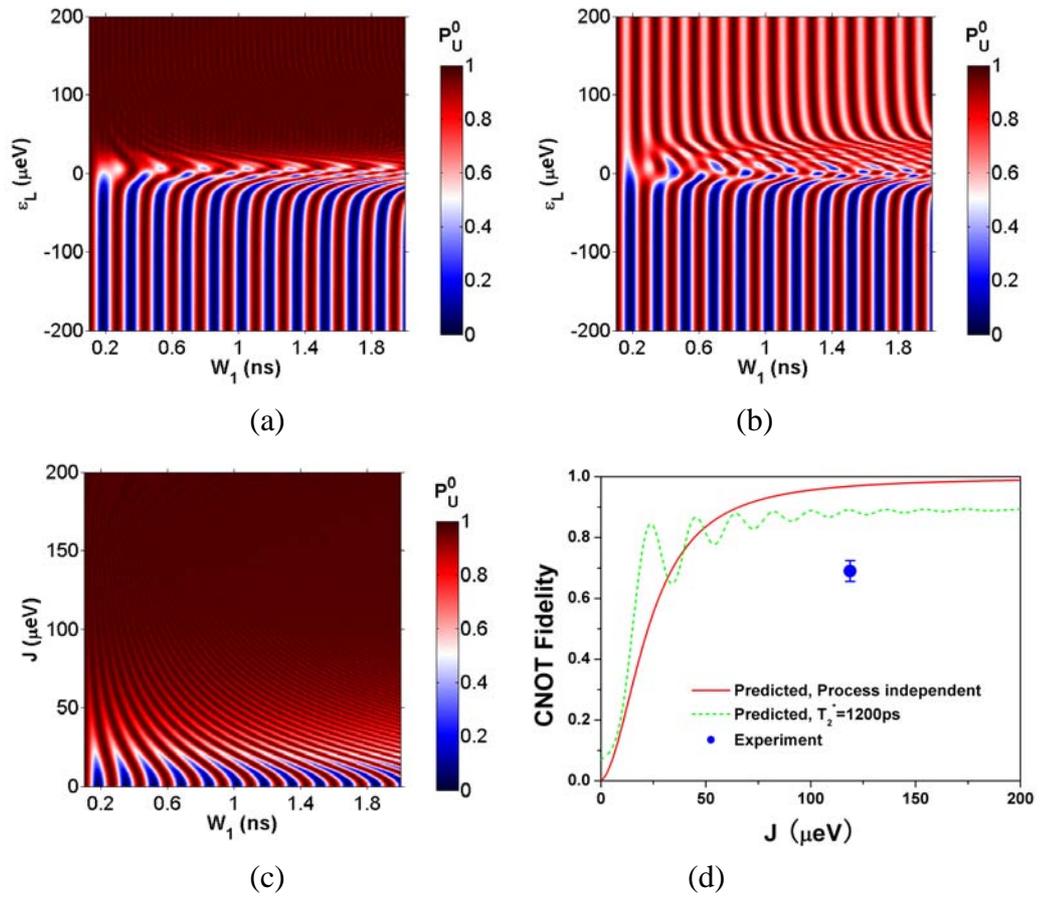

**Figure S2 Simulaition of CNOT Fidelity as a Function of *J***

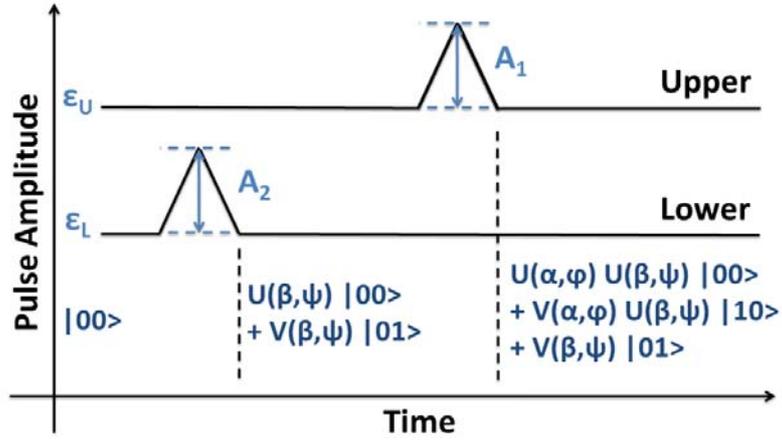

(a)

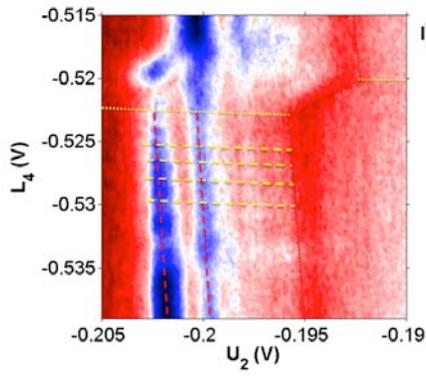

(b)

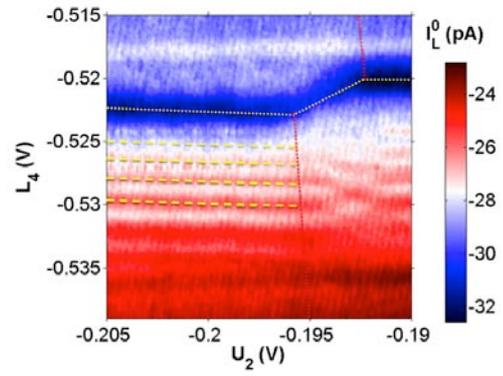

(c)

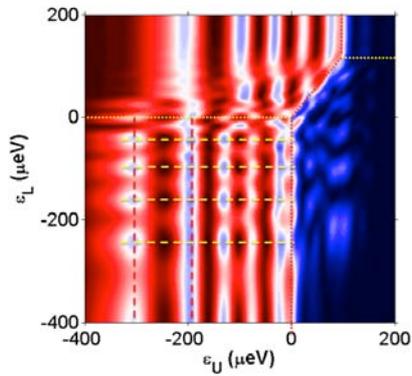

(d)

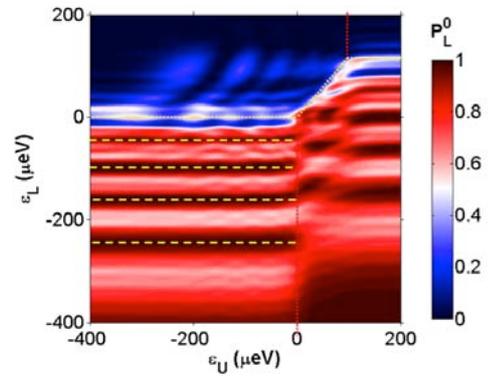

(e)

**Figure S3 Controlled-Universal-Rotations**

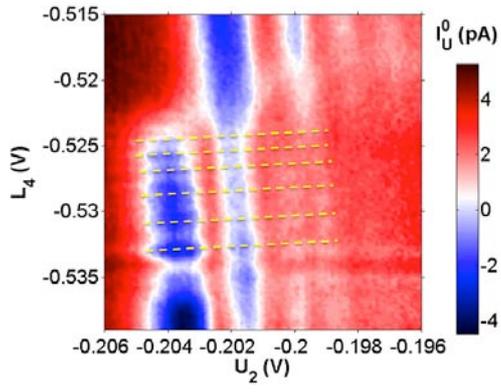
(a)
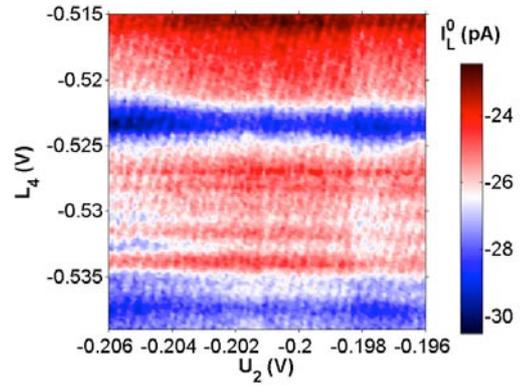
(b)
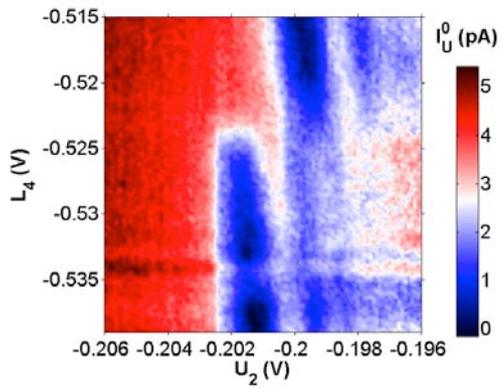
(c)
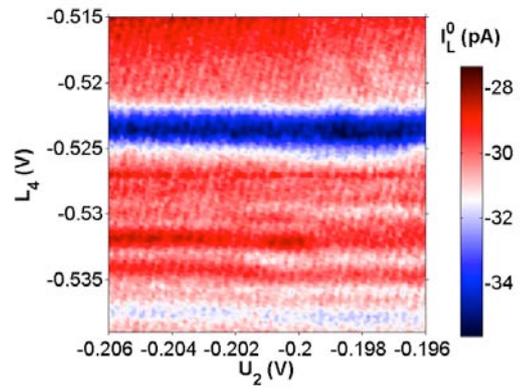
(d)
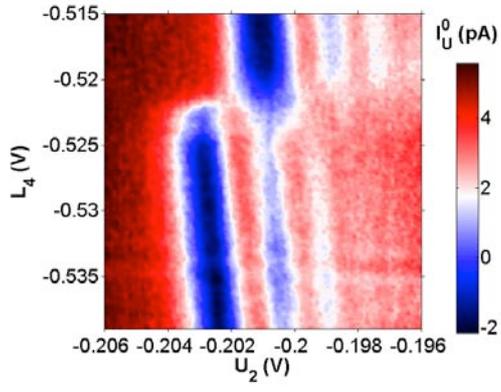
(e)
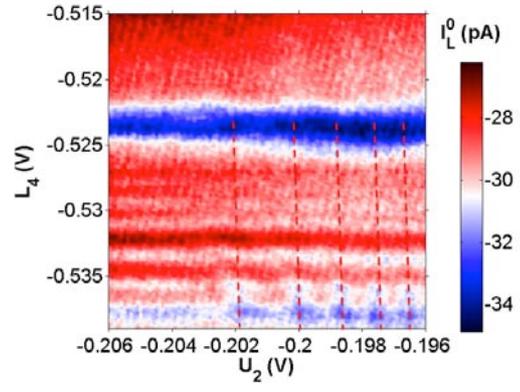
(f)

**Figure S4 Pulse Synchronization**